\documentclass[10pt,conference]{IEEEtran}

\usepackage{enumitem}

\usepackage{cite}
\usepackage[english]{babel}
\usepackage[font=small, textfont=bf, skip=0pt]{caption}
\usepackage{amsmath}
\usepackage{hyperref}
\usepackage{graphicx}
\usepackage{booktabs}
\usepackage{enumitem}
\usepackage{listings}
\usepackage{algorithm}
\usepackage{subcaption}
\usepackage{algpseudocode}
\usepackage{tikz}
\usepackage{longfbox}
\usepackage{algorithm}

\definecolor{mred}{rgb}{.80,.12,.30}
\definecolor{grey}{rgb}{0.5,0.5,0.5}
\definecolor{purple2}{rgb}{.75,0,.85}
\definecolor{pistachio}{rgb}{0.38, 0.57, 0.25}
\definecolor{steelblue}{rgb}{.20,.35,.90}

\newcommand{\theirs}[0]{\textsc{Proviso}}

\newcommand{\pre}[0]{\texttt{M\_pre}}
\newcommand{\method}[0]{\texttt{M}}

\input{code_highlight}
\title{Program Structure Aware Precondition Generation}

\author{
    \IEEEauthorblockN{Elizabeth Dinella}
    \IEEEauthorblockA{Bryn Mawr College\\
    edinella@brynmawr.edu}
    \and
    \IEEEauthorblockN{Shuvendu Lahiri}
    \IEEEauthorblockA{Microsoft Research\\
    shuvendu@microsoft.com}
    \and
    \IEEEauthorblockN{Mayur Naik}
    \IEEEauthorblockA{University of Pennsylvania\\
    mhnaik@seas.upenn.edu}
}

\begin{document}
%\begin{abstract}

   %  We present a general technique for inferring \textit{natural} preconditions from code. Our technique produces preconditions of high quality in terms of both correctness (modulo a test generator) and naturalness. Notably, the inferred preconditions are human readable and more easily understood by a language model. Our key insight is to leverage the structure of a method by employing a program reducer. We evaluate our approach showing it is effective in generating natural and correct preconditions. Lastly, as natural preconditions are desired by a language model, we instantiate 
   % our technique into a framework which can be applied at scale. We
  %  present a dataset of 17k Java (method, precondition) pairs obtained by applying our framework to \textcolor{red}{X} real-world projects. 

%\end{abstract}

\maketitle
We introduce a novel approach for inferring natural preconditions as a segment of code. Prior works, which generate preconditions from scratch through combinations of boolean predicates, often create preconditions which are difficult to read and comprehend. Our approach, instead, generates a precondition through a series of program transformations over the target method. Our evaluation shows that humans and Large Language Models can more easily reason over preconditions inferred using our approach. Lastly, we instantiate our technique into a framework which can be applied at scale. We present a dataset of ~18k Java (method, precondition) pairs obtained by applying our framework to 87 real-world projects.

%We introduce a novel approach for inferring \textit{natural} preconditions from code. Our technique produces preconditions of high quality in terms of both correctness (modulo a test generator) and naturalness. Prior works generate preconditions from scratch through combinations of boolean predicates, but fall short in readability and ease of comprehension. Our innovation lies in, instead, leveraging the structure of a target method as a seed to infer a precondition through program transformations. Our evaluation shows that humans can more easily reason over preconditions inferred using our approach. Lastly, we instantiate our technique into a framework which can be applied at scale. We present a dataset of $\sim$18k Java (method, precondition) pairs obtained by applying our framework to 87 real-world projects. We use this dataset to both evaluate our approach and draw useful insights for future research in precondition inference.

\section{Introduction}
\label{sec:intro}

\begin{figure*}
    %\centering
    \begin{minipage}{.4\linewidth}
    \begin{subfigure}{\linewidth}
    %\hspace{-2cm}
    \lstset{style=mystyle_lineno, xleftmargin=.0\textwidth, xrightmargin=0.0\textwidth, linewidth=0.2\linewidth}
\begin{lstlisting}[linewidth=8.8cm]
  public BigInteger[] DivideAndRemainder(BigInteger val) {
    if (val.m_sign == 0)
      throw new ArithmeticException("Division by zero error");
    BigInteger[] biggies = new BigInteger[2];
    if (m_sign == 0) {
      biggies[0] = Zero;
      biggies[1] = Zero;
    } else if (val.QuickPow2Check()) {
      int e = val.Abs().getBitLength() - 1;
      NetBigInteger quotient = Abs().ShiftRight(e);
      int[] remainder = LastNBits(e);
      biggies[0] = val.m_sign == m_sign ? quotient : quotient.Negate();
      biggies[1] = new BigInteger(m_sign, remainder, true);
    } else {
      int[] remainder = (int[]) m_magnitude.clone();
      int[] quotient = Divide(remainder, val.m_magnitude);
      biggies[0] = new BigInteger(m_sign * val.m_sign, quotient, true);
      biggies[1] = new BigInteger(m_sign, remainder, true);
    }
    return biggies;
  }

  public int getIntValue() {
    return m_sign == 0 ? 0 : 
            m_sign > 0 ?  m_magnitude[m_magnitude.length - 1]    
                       : -m_magnitude[m_magnitude.length - 1];
  }

\end{lstlisting}
    \label{fig:ex}
    \end{subfigure}
    \end{minipage}
    \hfill
    \begin{minipage}{.4\linewidth}
    %\centering
    \begin{subfigure}[p]{.7\linewidth}
    \vspace{0.1in}
    \lstset{style=mystyle_lineno, xleftmargin=.0\textwidth, xrightmargin=0.\textwidth, linewidth=0.0\linewidth}
\begin{lstlisting}[linewidth=5.5cm]
  boolean M_pre(BigInteger val) {
    if (val == null)
      return false;
    if (val.m_sign == 0)
      return false;
    return true;
  }
\end{lstlisting}
    \vspace{-0.05in}
    \centering
    \caption{Generated by our approach.}
    \vspace{0.15in}
    \label{fig:intro_ex_mine}
    \end{subfigure}
    \begin{subfigure}[p]{.7\linewidth}
    \lstset{style=mystyle_lineno, xleftmargin=.0\textwidth, xrightmargin=0.0\textwidth, linewidth=0.0\linewidth}
\begin{lstlisting}[linewidth=5.5cm]
  (not (val == null))
  and (
      (val.getIntValue() == val.m_sign and (
         (val.getIntValue() <= -1) or
         ((not (val.getIntValue() <= -1)) and 
             (not (val.getIntValue() <= 0))
         ))
      ) or
      (not (val.getIntValue() == val.m_sign))
  )
\end{lstlisting}
    \vspace{-0.05in}
    \caption{Generated by \theirs{}.}
    \label{fig:intro_ex_theirs}    
    \end{subfigure}
    \end{minipage}
    %\hfill
    %\begin{minipage}{.2\linewidth}
    %    \begin{subfigure}[p]{\linewidth}
    %    \vspace{0.1in}
    %\lstset{style=mystyle_lineno, xleftmargin=.0\textwidth, xrightmargin=0.\textwidth, linewidth=0.0\linewidth}
    %\lstinputlisting[linewidth=6.0cm]{./examples/DivideAndRemainder/DivideAndRemainder_context.java}
    %\vspace{-0.05in}
    %\caption{Generated by \textsc{Daikon}.}
    %\vspace{0.15in}
    %\label{fig:intro_ex_context}
    %\end{subfigure}
    %\end{minipage}

    \caption{\texttt{DivideAndRemainder} method and its corresponding preconditions.}
    \label{fig:mot_ex}
    %\vspace{-.7cm}
\end{figure*}

Methods in programming languages are partial functions that map a subset of inputs to an output value.
Unlike mathematical functions, a method is expected (according to the designer of the method) to be invoked on only a subset of all possible input values. 
Invoking a method on  illegal inputs can result in an exception, crash, or undefined behavior.
%, unless the method performs an explicit input validation. 
A boolean function which accepts all legal inputs and rejects all illegal inputs is a \textit{precondition}.

Preconditions serve a valuable purpose for both software engineering tools and programmers. Preconditions assist programmers in comprehension and correct usage of an API~\cite{api-san}. Reasoning about which inputs are legal vs.~illegal is critical to software correctness~\cite{cousot2, preconditions-classic}. Preconditions also enable effective application of automated testing~\cite{Randoop, dart, pex, EvoSuite}, static analysis~\cite{clang-sa, saturn} and verification~\cite{dafny}. In general, preconditions are essential for precise modular analysis. They play a significant role for both tools and programmers alike.

Various approaches have been proposed to automatically infer preconditions. Successful, pragmatic, approaches for generating preconditions are largely dynamic as static approaches~\cite{snugglebug, cousot, ESOP-2013-SeghirK} have difficulty scaling to real-world code bases.
%with constructs such as complex control flow, heap manipulations, interprocedural calls, and lengthy call chains. %don't know if we need this list. 
Dynamic approaches~\cite{data-driven, daikon, pre-mod-test} execute and monitor the program's variables and exit status (exceptional or non-exceptional). These approaches then generate a predicate through a search over logical features (e.g. \texttt{x > 0}), aiming to construct a precondition with accepts non-exceptional inputs and rejects ones which resulted in an exception. Constructing such a precondition has many challenges including selecting the relevant variables and constants, selecting applicable relationships between them (e.g. \texttt{<, ==, >=}), and structuring the relationships appropriately. State of the art approaches are successful in generating correct preconditions, but often select irrelevant variables, unintuitive relationships, and structure predicates in convoluted ways. Although such a generated precondition is correct in that it rejects illegal inputs and accepts legal ones, we show that the resulting predicates can become unnecessarily complex, difficult to comprehend, and ultimately, unnatural. 

%A precondition which accepts legal inputs and rejects illegal ones is not unique and there can be multiple "correct" preconditions on the given input space with some being more natural and easy to comprehend than others. For example, the predicate x > 0 is equivalent to (not x < -1).  

Natural preconditions are desirable for a variety of reasons.
Firstly, they are easier for programmers to consume and comprehend. 
%Human programmers, who are the creators and maintainers of software systems, likely find it easier to work with preconditions that are phrased in similar language. 
Natural preconditions are better suited for human-in-the-loop settings such as integrated development environments and interactive program verifiers. In general, natural preconditions are easier to understand and therefore easier to specify and refine, contributing to a more effective and accurate development or analysis~\cite{crowdsource-preconditions}. Likewise, natural preconditions phrased in a similar language to the original method are likely to be preferred by statistical models for code that increasingly underlie software engineering tools~\cite{naturalness-of-software}. Leveraging natural preconditions aligns well with the evolving landscape of software engineering tools and statistical modeling to enhance the efficacy of human-involved processes~\cite{sutton-daikon, are-my-invariants-valid, c2s}.

In this paper, we propose a dynamic approach for inferring natural and correct preconditions. In contrast to existing approaches, we infer a precondition by performing transformations to the target method. Rather than constructing preconditions from scratch through feature combination, we leverage the structure of the method as a seed for inference. Our approach actively queries a test generator, iteratively finding crashing inputs and transforming the precondition to effectively guard against them. Our approach converges on a precondition which is in the form of a boolean returning function rather than a predicate. In this paper we detail our proposed technique and discuss the challenges involved in developing an algorithm to infer preconditions through program transformation.

%1) How to keep only segments of code related to the precondition
%2) How to guard against exceptions that aren't explicitly in the source 
%3) How to deal with interprocedural crashes. 

%Through this, we arrive upon an algorithm which iterates between input generation and refined program transformation, ultimately converging on a correct and natural precondition. 

%We introduce a pragmatic and efficient framework that tackles the intricacies of real-world constructs found in modern programming languages including complex control flow, heap manipulations, interprocedural calls, and lengthy call chains. 

We evaluate our approach in terms of both naturalness and correctness by performing a comparative study to prior work. We find that our preconditions are comparably correct yet preferred by both humans and Large Language Models (LLMs). By conducting a user study, we find that human participants were able to more accurately reason over our preconditions in a shorter time span versus the state-of-the-art approach. Consumers of our preconditions completed reasoning tasks more accurately (88.84\%) with an average total duration of 150 seconds. In contrast, consumers of the preconditions inferred by prior work took longer (205 seconds) to finish the study and answered with lower accuracy (75.71\%). By using inferred preconditions in few-shot prompts to GPT-4, we find that our preconditions improve output of the LLM (71.78\% accuracy vs 53.84\% accuracy zero-shot) while preconditions inferred by prior work led to marginal improvement (58.97\% accuracy).

Finally, we present the implementation of our approach as a tool and demonstrate its application on a significant scale across 87 real-world Java projects. This yields $\sim$18k (method, precondition) pairs, obtaining the first large-scale dataset of preconditions~\cite{c2s, crowdsource-preconditions}. This dataset not only underscores the viability of applying our tool at scale, but also provides a target for future real-world training and evaluation. In this paper, we use this dataset to conduct a large-scale evaluation of the design decisions of our approach and present characteristics of the dataset for future research.
%\TODO{Add something about the implication that ours improves GPT outputs few shot}.

%which in turn offers valuable insights to guide future research efforts in the field of precondition inference.\TODO{fix this paragraph I really don't like it.}
 
\noindent In summary, our work makes the following contributions:
\begin{enumerate}
\item We introduce a novel methodology for inferring natural preconditions through program transformation. 

\item We introduce a new representation for preconditions as a boolean returning function.

\item We present an instantiation of our approach as a tool which can be applied at scale. 

\item We perform a comparative evaluation to state-of-the-art precondition generation approaches and show that ours are more easily reasoned over by humans.

\item  We evaluate the effectiveness of the inferred preconditions as few-shot examples to GPT-4 and find that our preconditions are more effective in improving accuracy.

%GPT-4's affinity 
%\TODO{and large language models} through a user study. \TODO{mention Daikon experiment too}.
%\item \TODO{GPT experiment????}

\item We curate a dataset of $\sim$18k samples and present an exploratory study of its characteristics. 

%\item We perform a large-scale evaluation of the decisions made to design our approach, showing they contribute to higher quality preconditions. \TODO{Do I really want to highlight this??}
\end{enumerate} 
Our tool and dataset are included anonymously as supplementary material in the submission. We plan to open-source and publicly release both upon publication.

\section{Motivating Examples}
\label{sec:mot_ex}
% General note: I prefer "method implementation" over plain "method", although it can get wordy and so it is OK to shorten when it is obvious from contenxt.

In this section, we motivate our approach using the example in Figure~\ref{fig:mot_ex}. Consider a Java method \texttt{DivideAndRemainder}, for which we aim to infer a precondition. The target method implementation for dividing arbitrary-precision integers is shown on the left of the figure. The \texttt{BigInteger} object is stored with a field \texttt{m\_sign} (0, -1, or 1) and an \texttt{m\_magnitude} storing the integer value. The method throws an \texttt{ArithmeticException} when the denominator is zero, which is captured by the check (\texttt{val.m\_sign == 0}) on line 2. Additionally, the method throws a \texttt{NullPointerException} upon execution of line 2 when \texttt{val} is null. A correct precondition for this method is one that does not allow inputs where the denominator argument {\tt val} is either the \texttt{BigInteger} zero or \texttt{null} and accepts all other inputs: (\texttt{val != null and val.m\_sign != 0}).

The popular invariant detector \textsc{Daikon} generates a precondition that, although correct, contains 43 clauses and references 5 irrelevant fields. For instance, the precondition contains the clause: \texttt{this.m\_magnitude.length \% val.m\_numBits == 0}. Although true in all non-crashing runs, these variables and their relationships are unrelated to the precondition. \textsc{Proviso} improves upon Daikon by generating the precondition in Figure~\ref{fig:intro_ex_theirs} through the ID3 classification algorithm, selecting features with the largest information gain. Although correct, \textsc{Proviso}'s generated precondition references the irrelevant function \texttt{getIntValue}. The generated precondition guards against the case when \texttt{val} is zero by comparing the output of \texttt{getIntValue} to \texttt{val.m\_sign}. In the case where the parameter \texttt{val} is zero, the output of \texttt{getIntValue} and \texttt{val.m\_sign} are coincidentally equal (as in the case of 1 and -1). The inequalities in lines 4-7 further specify that the output of \texttt{getIntValue} must be equal to zero. By combining predicates that may be incidentally correct, the resulting precondition becomes unnecessarily complex, containing irrelevant variables and relationships.

In contrast, we find that leveraging the program structure and performing program transformations often leads to more natural preconditions. Our approach is able to generate the precondition as a segment of code shown in Figure \ref{fig:intro_ex_mine}. Our precondition only references relevant variables and is structured in a readable format.

An approach to generate preconditions through program transformations to the target method gives rise to two subproblems: (1) guarding against all crashes (2) removing code irrelevant to the precondition. In Figure~\ref{fig:intro_ex_mine}, guarding against crashes involved transforming the explicitly thrown \texttt{ArithmeticException} into a \texttt{return false} and inserting the nullness check prior to line 2. Lines 4-24 (in the target method) are unrelated to the precondition and were removed. %We also include an example, Figure~\ref{fig:16626}, from our dataset to illustrate the complexity of real-world preconditions. In this example, the subproblems amounted to (1) inserting three nullness checks corresponding to the field accesses marked in red  (Figure~\ref{fig:16626:mut}) and (2) removing the initialization and modification of \texttt{votes}. 
In Section~\ref{sec:formulation} we discuss the challenges of these two subproblems as well as a discussion of our design decisions in Section~\ref{sec:technique}.

\section{Problem Formulation}
\label{sec:formulation}

\begin{figure*}
\centering
\includegraphics[width=.88\textwidth]{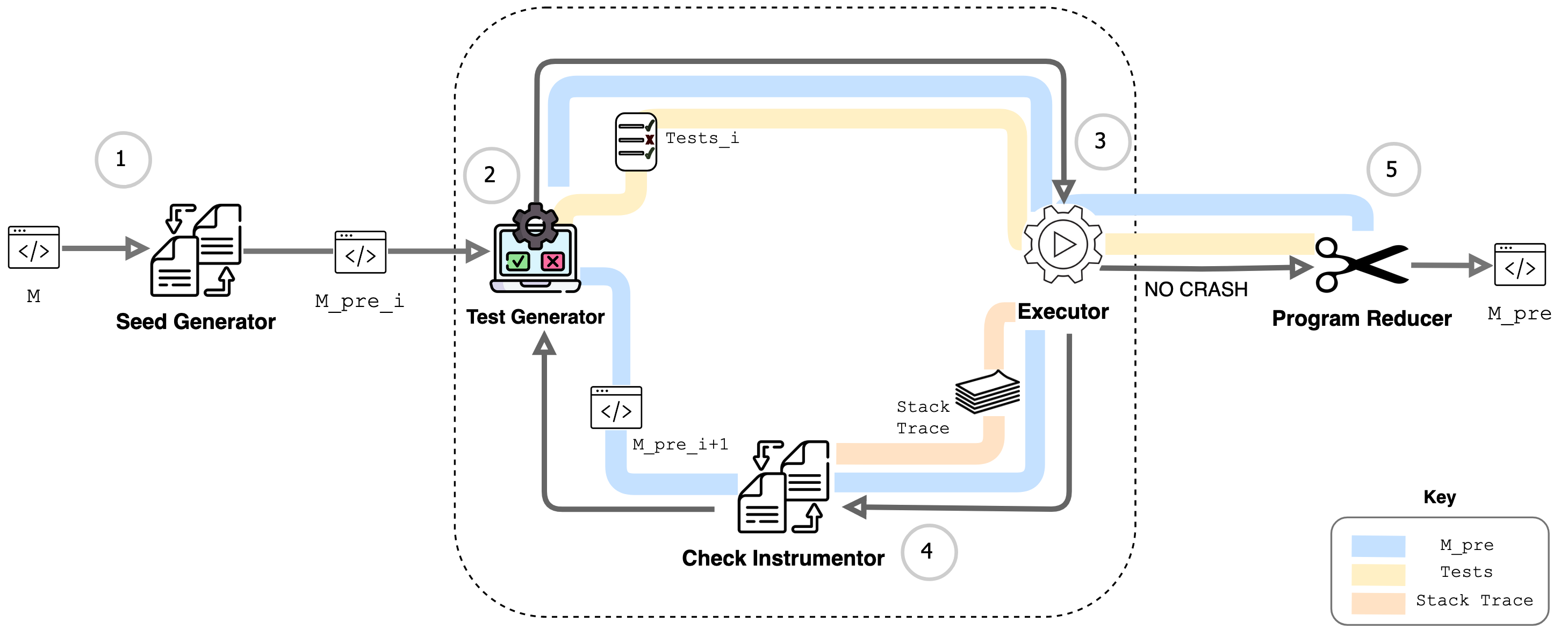}
        \caption{Overall architecture of our approach.}
        \label{fig:technique}
\end{figure*}

In this section, we formulate the problem of inferring preconditions through program transformation. Given a deterministic method \method, we infer a precondition, \pre{}. 
We aim to generate a precondition which is \emph{safe} and \emph{maximal}. A safe \pre{} is one that rejects any illegal input while a maximal \pre{} is one that accepts any legal input. 
%Following the definition of correctness from \cite{pre-mod-test}, 

Given an exhaustive tester \texttt{T} which generates a set of input environments \texttt{E}, we aim to infer an \pre{} with the following correctness properties for every $e \in \texttt{E}$:
\begin{quote}
$\pre{}(e) \rightarrow True$ iff \method{} exits normally on $e$ and \\ 
$\pre{}(e) \rightarrow False$ iff \method{} throws an exception on $e$.
\end{quote}

\subsection{Subproblems}
A dynamic algorithm for inferring preconditions through program transformation comes with unique challenges.

%Deriving an algorithm for \pre{} inference, through a conjunction of program transformation and test generation, comes with challenges.

\paragraph{Guarding Against Crashes}
In order for \pre{} to be safe, it must include checks to guard against all feasible crashes. To do so, one must carefully design program transformations to insert guards which precisely reject illegal inputs. In Section~\ref{sec:technique} we discuss design decisions in various stages of our algorithm to ensure precise guards which do not inadvertently introduce error by rejecting legal inputs. This proves to be challenging as it is not obvious how to design program transformations to guard against a variety of crash types. For instance, it is unclear how to insert source checks for interprocedural crashes. Secondly, we show that one must consider when $T$ should be invoked to generate $E$. In Section~\ref{sec:technique} we detail our algorithm which interleaves program transformation and test generation in an iterative process until convergence. We motivate the need for active queries to $T$ in Section~\ref{sec:dataset}. 

%One obvious algorithm is \textit{exhaustive up-front transformation} which systematically inserts checks, prior to invoking $T$. Presumably, it is possible for a static exploration to find all JVM level exception checks which can then be translated to exhaustive source level checks. We also present an alternate \textit{single step non-iterative} algorithm. This algorithm begins by invoking $T$ once, and insert checks for crashing inputs to create \pre{} in a non-iterative process. However, neither of these approaches are effective in practice: the former does not guarantee correctness, whereas the latter can result in a poor-quality test suite \texttt{E}. 
%We illustrate these issues in Section \ref{sec:technique:naive}  %and propose to address them by inserting exception checks both {\it iteratively} and {\it dynamically}. In particular, 
 %In section~ we show that inserting exception checks both iteratively and dynamically is necessary for correctness.  

\paragraph{Removing Irrelevant Computation}
Upon lifting exception checks, our transformed precondition is correct modulo the test generator. However, it contains many program segments which are unrelated to the precondition. For illustration, consider the example from Figure~\ref{fig:mot_ex}. Lines 4-24 are unrelated to the precondition. We present a dual rationale for eliminating these segments. Firstly, we emphasize the importance of removal for naturalness. To ensure that \pre{} is conducive to human reasoning it should not include program segments which are unrelated to the precondition and incur additional cognitive load. Secondly, these extraneous segments might encompass statements or expressions with side effects. A pure (i.e., side-effect free) precondition is desirable in both ease of understanding and in tool usage. In tool usage, a precondition with side effects must be run in a sandbox environment as to preserve the state of the program being analyzed. The challenge of maintaining purity is unique to our approach. Prior works have the benefit of ensuring purity by construction. That is, their precondition features are designed to only include observer methods which do not change the program state. In Section~\ref{sec:technique} we describe our solution of dynamic program reduction which we employ for efficient removal of irrelevant computation. 

\section{Technique}
\label{sec:technique}

 \begin{figure}
    \setlength{\belowcaptionskip}{1em}
    \begin{subfigure}[]{.45\linewidth}
    \lstset{style=mystyle, xleftmargin=.0\textwidth, xrightmargin=0.0\textwidth}
\begin{lstlisting}[linewidth=3.8cm]
BigInteger Sqrt() {
   if (m_sign < 0)
      throw new Exception();


   return Round(Sqrt(this));

}

\end{lstlisting}
    \caption{Target method \method{}.}
    \label{fig:seed_gen:before}
    \end{subfigure}
    \begin{subfigure}[]{.45\linewidth}
    \lstset{style=mystyle, xleftmargin=.1\textwidth, xrightmargin=0.0\textwidth}
\begin{lstlisting}[linewidth=4.5cm]
boolean Sqrt_pre() {
   if (m_sign < 0)
        return false;
 
   BigInteger var_a = Sqrt(this);
   Round(var_a);
   return true;
}
\end{lstlisting}
    \caption{\method{} after seed transformation.}
    \label{fig:seed_gen:after}
    \end{subfigure}
    %\vspace{1em}
    \caption{Illustration of seed generation.}
    \label{fig:seed_gen}
\end{figure}
%\vspace{1em}

In this section, we present the details of our general approach as well as our instantiated framework. We describe the phases involved in inferring a precondition through program transformation. We address sub-problems (1) guarding against crashes and (2) removing code irrelevant to the precondition. Our overall technique to generate a precondition through program transformation is illustrated in Figure~\ref{fig:technique}. The initial source transformation phase, described in Section~\ref{sec:technique:init}, creates a seed for inference. This up-front step transforms \method{} into a boolean returning function \texttt{M\_pre\_i} which resembles the original method structure. Next, program transformations are iteratively applied to the seed to guard against crashes. This phase of the approach, described in Section~\ref{sec:technique:sp1}, tackles sub-problem (1) by adding explicit checks in the source of \texttt{M\_pre\_i} upon actively querying a test generator. At a high level, our approach discovers crashes by actively generating regression tests and executing them. Upon executing an exceptional input, the stack trace provides us with the crash type and line number. This allows us to instrument \texttt{M\_pre\_i} with precise checks that guard against a particular illegal input. We present a set of six transformations which are selected and applied based on given crash type. For instance, a \texttt{NullPointerException} should be guarded against by inserting a nullness check. Lastly, our technique tackles sub-problem (2) by leveraging a syntax-guided program reducer, described in Section~\ref{sec:technique:sp2}. The final transformed program \pre{} contains checks to guard against all found crashes and does not contain irrelevant computation. Ultimately, this results in a natural and correct precondition. %In Section~\ref{sec:technique:naive} we motivate the need for an iterative transformation process by illustrating the incorrectness incurred through alternate approaches presented in Section~\ref{sec:formulation}. 

%The last phase, described in Section~\ref{sec:technique:sp2}, tackles sub-problem (2) by leveraging a syntax-guided program reducer. 

%HERE I SHOULD PERHAPS INSERT THE ADD EXAMPLE?
%At a high level, our approach works in multiple 

%Each phase works in synergy benefiting from design choices made in other phases. 

%The seed generation phase localizes expressions for precise exception check insertion and syntax-guided reduction. 

%The second phase, benefits from the first, and relies on the reducer's (third phase) properties to remove conservative checks. The reducer benefits from the previous phases as it receives a \pre{} with localized exception checks with a reducer-aware structure. 

\subsection{Seed Generation}
\label{sec:technique:init}
Our overall technique begins by creating a seed through an up-front source transformation on \method{}. The seed retains the structure of the target method, but has small differences to satisfy our problem formulation, requiring \pre{} to be a non-exceptional, boolean returning function. Our seed generation process also makes semantics preserving transformations for precise exception check insertion and syntax-guided reduction in later stages. In particular, when inserting checks to guard against illegal inputs, it is important to carefully select the correct source location to maintain maximality and readability. An imprecisely placed check could result in rejecting a legal input. 

Figure~\ref{fig:seed_gen} illustrates the seed source transformation on an example method \texttt{Sqrt}. The source transformation is designed such that the seed has the following desirable qualities: 

%\begin{enumerate}
\paragraph{The \pre{} must be boolean returning:} 
\noindent We transform ${\tt M}$ to be a boolean returning function by modifying the method signature and adding the statement \texttt{return true} at all exit points. If \method{} returns a type other than boolean, we maintain the original expression return value by lifting it to an expression statement prior to the newly inserted return statement. This is essential as the return expression in~\method{} itself may be exceptional. 
%\begin{enumerate}[noitemsep]
\paragraph{The \pre{} must be non-exceptional:}
\noindent Our problem formulation requires~\pre{} to either return true or false, implying that it exits normally on all inputs. At times, programmers will explicitly guard against precondition violations, by explicitly throwing an exception. For example, in Figure~\ref{fig:seed_gen:before} an \texttt{Exception} is thrown on a zero input. To ensure the~\pre{} exits normally, we replace any \texttt{throw} expression with a \texttt{return false} indicating an illegal input. The target method (and seed) may contain feasible crashes which are later guarded against in our iterative approach described in Section~\ref{sec:technique:sp1} and illustrated in the dotted box in Figure~\ref{fig:technique}.
\paragraph{The \pre{} localizes crashes:}
\noindent The seed generation phase transforms \method{} such that any checks inserted in a later phase precisely localize the crash. To illustrate this, consider Figure~\ref{fig:norm} in which a \texttt{NullPointerException} may potentially occur upon the field access on \texttt{a} in the loop condition. Inserting a check at P1 (prior to the loop declaration) does not guard against exceptions which may occur during loop execution. Since \texttt{taz()} may return \texttt{null}, a guard prior to entry of the loop is not sufficient. During seed generation, we normalize potentially exceptional expressions in loop conditions such that guards can be precisely inserted in the next phase of inference. Normalization for \texttt{for} loop conditions follow similarly.

\begin{figure}
    \setlength{\belowcaptionskip}{1em}
    \begin{subfigure}[]{.4\linewidth}
    \lstset{style=mystyle, xleftmargin=0.0\textwidth, xrightmargin=0.0\textwidth}
\begin{lstlisting}[linewidth=3.8cm]
void foo(Object a) {

    int i = 0;
    //P1
    while (i < a.bar) 
        //taz may return null
        a = taz(); 

        i += 1;

    } 
    
}
\end{lstlisting}
    %\caption{\method{} with an exceptional loop condition.}
    \label{fig:norm:init}
    \end{subfigure}
    %\begin{subfigure}[]{.45\linewidth}
    %\lstset{style=mystyle, xleftmargin=0.1\textwidth, xrightmargin=0.0\textwidth}
    %\lstinputlisting[linewidth=5.1cm]{./examples/Seed/intra_bad.java}
    %\caption{Incorrect check insertion.}
    %\label{fig:norm:bad}
    %\end{subfigure}
    \begin{subfigure}[]{.45\linewidth}
    \lstset{style=mystyle, xleftmargin=0.1\textwidth, xrightmargin=0.0\textwidth}
\begin{lstlisting}[linewidth=4.2cm]
boolean foo_pre(Object a) {
    int i = 0;
    while (true) 
        //P1
        if (i < a.bar) {
            a = taz(); 
        } else {
            break;
        }
        i += 1;
    } 
    return true; 
}
\end{lstlisting}
    %\caption{\method{} after seed transformation with normalization.}
    \label{fig:norm:seed}
    \end{subfigure}
    %\begin{subfigure}[]{.45\linewidth}
    %\lstset{style=mystyle, xleftmargin=0.1\textwidth, xrightmargin=0.0\textwidth}
    %\lstinputlisting[linewidth=5.1cm]{./examples/Seed/intra_good.java}
    %\caption{Correct check insertion.}
    %\label{fig:norm:good}
    %\end{subfigure}
    \caption{Illustration of loop normalization in seed generation for precise check insertion. After applying the normalization (right), a nullness check at P1 would correctly guard against crashes.}
    \label{fig:norm}

\vspace{1em}
\end{figure}

Another major challenge in generating a precondition through program transformation is in regards to interprocedural crashes. In Section~\ref{sec:technique:sp1} we describe and motivate our approach to tackle this through a predefined program transformation. In short, this transformation amounts to wrapping an exceptional call in a \texttt{try-catch} block. As a product of this transformation we perform \textit{call normalization} on our seed. The process of call normalization is essential for crash localization during the next phase of transformation. The normalization transformation lifts each call to its own source line. Call normalization ultimately results in a more readable precondition. This is illustrated in Figure~\ref{fig:call_norm}. Suppose \texttt{Sqrt} throws a general type \texttt{Exception} on an input. Without call normalization, the final ~\pre{} is shown in Figure~\ref{fig:call_norm:no}. It is not clear whether the exception is occurring in the method \texttt{Sqrt} or the method \texttt{Round}. On the other hand, performing call normalization (Figure~\ref{fig:call_norm:yes}) localizes the exception in \texttt{Sqrt}. By localizing the crash we reduce the cognitive load of interprocedural inspection of the exceptional callee.

We perform this initial source transformation up-front to create a seed \pre{} for our forthcoming phases. All transformations beyond the boolean returning behavior are semantic preserving transformations. %\footnote{We do not currently support some Java constructs (do-while loops, switch statements, synchronized statements, and labelled statements) in our framework due to engineering work.}
%We motivate our design decision for explicitly guarding against interprocedural crashes through \texttt{try-catch} blocks in Section~\ref{sec:technique:sp1}.

%\end{enumerate}

\begin{figure}
\centering
    \setlength{\belowcaptionskip}{1em}
    \begin{subfigure}[]{.45\linewidth}
    \lstset{style=mystyle, xleftmargin=0.0\textwidth, xrightmargin=0.0\textwidth}
\begin{lstlisting}[linewidth=3.8cm]
public boolean Sqrt_pre() {
    if (m_sign < 0) 
        return false;
    try {  
        Round(Sqrt(this));
    } catch (Exception e) {
        return false;
    }
    return true;
}
\end{lstlisting}
    \caption{Without call normalization.}
    \label{fig:call_norm:no}
    \end{subfigure}
    \begin{subfigure}[]{.45\linewidth}
    \lstset{style=mystyle, xleftmargin=.1\textwidth, xrightmargin=0.0\textwidth}
\begin{lstlisting}[linewidth=4.0cm]
public boolean Sqrt_pre() {
    if (m_sign < 0) 
        return false;
    try {  
        Sqrt(this);
    } catch (Exception e) {
        return false;
    }
    return true;
}
\end{lstlisting}
    \caption{With call normalization.}
    \label{fig:call_norm:yes}
    \end{subfigure}
    %\vspace{1em}
    \caption{Illustration of call normalization in seed generation for precise guarding against interprocedural crashes.}
    \label{fig:call_norm}
\end{figure}
%\vspace{2em}
%\end{enumerate}

\algnewcommand\algorithmicswitch{\textbf{switch}}
\algnewcommand\algorithmiccase{\textbf{case}}

\algdef{SE}[SWITCH]{Switch}{EndSwitch}[1]{\algorithmicswitch\ #1\ \algorithmicdo}{\algorithmicend\ \algorithmicswitch}%
\algdef{SE}[CASE]{Case}{EndCase}[1]{\algorithmiccase\ #1}{\algorithmicend\ \algorithmiccase}%
\algtext*{EndSwitch}%
\algtext*{EndCase}%

\begin{algorithm*}
{\footnotesize
\caption{Transformation to replace a crashing program fragment on Line L with a new program fragment that guards the crashing exception type.}
\label{algo:check_insertion}
\begin{algorithmic}[1]
\Procedure{CheckInstrumentorAlgorithm}{Exception Type T, Line L, bool isInCallee}
\State S $\gets$ statement on line L %\Comment{seed generation ensured that S is in a nice form}
\State P $\gets$ S

\If{isInCallee} \Comment{Interprocedural Exception}
        \State P $\gets$ try \{ S \} catch (T exc) \{ return false; \}  %\Comment{S is of the form \texttt{func\_expr(arg\_expr\_list)} due to call normalization.}
        \State \texttt{return;}
\EndIf

\State{Let (\texttt{expr\_1}, ..., \texttt{expr\_n}) be the list of all expressions in S in the normal order of execution} %(i.e. postorder traversal)}

\For {$i = 1 : n $}
    \Switch{T} 
        \Case{\texttt{java.lang.NullPointerException}}
        
            \If {expr\_i is a field access expression of the form \texttt{object\_expr.field}}
                  \State  P $\gets$ \texttt{{ if (object\_expr == null) return false; }} P
            \EndIf
           
            \If {expr\_i is an array access expression of the form \texttt{array\_expr[index\_expr]}}
                  \State  P $\gets$ \texttt{{ if (array\_expr == null) return false; }} P
            \EndIf
       \EndCase

        \Case{\texttt{java.lang.ArrayIndexOutOfBoundsException}}
            \If {expr\_i is an array access expression of the form \texttt{array\_expr[index\_expr]}}
                \State  P $\gets$ \texttt{{ if (index\_expr < 0 || index\_expr >= array\_expr.length) return false; }} P
            \EndIf
        \EndCase

        \Case{\texttt{java.lang.ClassCastException}}
            \If {expr\_i is a class cast expression of the form \texttt{(cast\_type) expr\_to\_cast}}
                  \State  P $\gets$ \texttt{{ if !(expr\_to\_cast instanceof cast\_type) return false; }} P
            \EndIf
        \EndCase

    \Case{\texttt{java.lang.NegativeArraySizeException}}
         \If {expr\_i is an array creation expression of the form \texttt{new array\_expr[index\_expr]}}
                \State  P $\gets$ \texttt{{ if (index\_expr < 0) return false;}} P
        \EndIf
    \EndCase

    \Case{\texttt{java.lang.ArithmeticException}}
        \If {expr\_i is a binary operator \texttt{DIVIDE} expression of the form \texttt{numer/denom}}
            \State  P $\gets$ \texttt{{ if (denom == 0) return false;}} P
        \EndIf
  \EndCase

\EndSwitch 
       
\EndFor
\EndProcedure

\end{algorithmic}
}
\end{algorithm*}

\subsection{Exception Check Insertion}
\label{sec:technique:sp1}
In this section we describe our solution to sub-problem (1), guarding against crashes. We do this by iteratively interleaving program transformation and test generation. This phase of our approach is shown in the dotted box in Figure~\ref{fig:technique}. It begins by operating on the seed program $\pre{}\_i$. In order to discover crashes, we invoke the test generator to obtain a set of tests, ${\tt Tests}\_i$. We then execute the tests and observe the stack trace(s) of all crashing inputs. If crashing inputs do indeed exist in ${\tt Tests}\_i$, then the $\pre{}\_i$ should be transformed such that any crash instead exits normally with a return value of ~\texttt{false}. To do so, we invoke the check instrumentor which performs program transformations to satisfy this formulation. The check instrumentor modifies the semantics of $\pre{}\_i$ to produce an $\pre{}\_{i+1}$, by adding checks (false returning if-statement guards) prior to any crash found by the test generator. After transformation, we again invoke the test generator on the newly produced $\pre{}\_i+1$ to yield a new set of tests ${\tt Tests}\_i+1$. If this new set of tests contains new crashing inputs which are not already guarded against, we continue in the loop and invoke the check instrumentor to guard against any new crashes. We continue this iterative process until convergence. That is, when we no longer observe crashes during execution of the current ${\tt Tests}\_i$. Upon convergence, we are guaranteed to have a safe and maximal precondition, modulo the test generator. 
%In Section~\ref{sec:technique:naive} we illustrate the need for such an iterative process to produce correct and high quality preconditions. 
Inserting exception checks both iteratively and dynamically is necessary to guarantee correctness. Furthermore, inserting exception checks on demand creates higher quality preconditions by driving the test generator to progressively higher coverage inputs. This allows for the discovery of additional crashes and insertion of corresponding explicit checks. We quantify the number of iterations before convergence in section~\ref{sec:dataset}.

%An algorithm which conservatively guards against any potential exception through static checks risks inserting an imprecise check. Discovering concrete crashing inputs allows us to utilize the stack trace. 

%and then calling the reducer is insufficent and can lead to incorrect preconditions. 

%In this section, we begin by describing our transformation types then discuss the iterative process shown in the dotted box in Figure~\ref{fig:technique}. 
%When $T$ is applied to the current $M\_pre\_i$, EvoSuite may find crashing and non-crashing inputs. According to our problem formulation an exception should not be thrown on an illegal input, but instead return false. To tackle this, our technique executes the EvoSuite generated tests to obtain a stack trace. Any exception must be handled by a check inserted through program instrumentation so it returns false rather than crashing. In the case where $T$ does indeed find an exceptional $e \in E$, we use AST transformations to insert an explicit check.

%\TODO{need to say somewhere we can't just glean out static checks.}

\paragraph{\textbf{Check Instrumentor}} Here, we describe the process for inserting \texttt{false} returning guards prior to crashes found by the test generator. These are inserted such that the \pre{} will exit normally on an illegal input rather than throwing an exception. The check instrumentor parses a stack trace produced from the execution of the current tests ${\tt Tests}\_i$. The stack trace provides a crash type and location, which allows us to make precise AST transformations. By only guarding against the given crash type at the given location, we maintain maximality and do not reject any legal inputs. We categorize and design our program transformations in a data-driven fashion. We define six transformations which guard against 99\% of the crashes the test generator found on 87 real-world Java projects. Our technique performs AST transformations according to the Algorithm~\ref{algo:check_insertion}. This algorithm works in synergy with our seed generation as it expects localized statements to match the given line number. The first transformation, on Lines 4-7, is performed when the crash occurs in a callee of \method{}. This amounts to wrapping the crashing call in a \texttt{try-catch} block. The seed generation call normalization guarantees that $S$ is a single method call to ensure precise localization. Lines 8-34 handle the cases where the crash occurs directly in the target method. Here, the statement on Line $L$ can contain multiple child expressions which are iterated over in order of execution. Our algorithm similarly transforms the program segment based on crash and expression types. For instance, given an \texttt{ArrayOutOfBoundsException} crash type,  the transformation on line 20 is performed for each \texttt{ArrayIndexExpression} on line $L$. If multiple index expressions are present as children of $S$, guards are conservatively added for each. Working in synergy with the reducer, if the checks are unnecessary, they will be removed in a later stage of the technique. The other four transformation types follow similarly.

\paragraph{A Note on Interprocedural Wraps} It is not obvious how to add explicit checks for crashes that occur within callees in~\method{}. One possibility is to inline any calls. We experimented with inlining method calls, but found it to lead to lower quality or equivalent preconditions for two reasons. Firstly, inlining can quickly hit a library or framework boundry where the source code is not included. Secondly, even when all source is available, inlining often leads to potentially complex precondition logic due to long call chains with large method bodies. Through a manual inspection of 10 random samples from our dataset, we found that only one sample benefited from inlining. Furthermore,  we found that over 85\% of the exceptional callees in our dataset are at a source boundary and cannot be inlined. For these reasons, we choose to handle exceptions in the callee by wrapping the crashing call in a \texttt{try-catch} block. %Lastly, we note that our catch expression captures the Exception type seen in the stack trace rather than the general Exception type in order to further localize crashes. 

%\TODO{Reduce font size a bit as this looks ugly.  We could also be a bit more mathematical and put this into a figure instead. For instance, I don't think the scope of "..." is accurate: E.g. in Rule (1), how early do I introduce the "if (a == null) return false"? Ditto for other rules.}
%\begin{enumerate}
%\item \texttt{java.lang.NullPointerException} \\
%$\texttt{...\textbf{a.foo}...} \rightarrow \texttt{if (a == null) return false;...\textbf{a.foo}...}$

%\item \texttt{java.lang.ArrayIndexOutOfBoundsException} \\
%$\texttt{...\textbf{a[i]}...} \rightarrow \texttt{if (\textbf{i} < 0 || \textbf{i}  >= \textbf{a}.length) return false;...\textbf{a[i]}...}$

%\item \texttt{java.lang.ClassCastException} \\
%$\texttt{...\textbf{(T) a}...} \rightarrow \texttt{if (!\textbf{a} instanceof \textbf{T}) return false; ...\textbf{(T) a}...}$

%\item \texttt{java.lang.NegativeArraySizeException} \\
%$\texttt{...\textbf{new T[i]}...} \rightarrow \texttt{if (\textbf{i} < 0) return false; ...\textbf{new T[i]}...}$

%\item \texttt{java.lang.ArithmeticException}\\
%$\texttt{...\textbf{a/b}...} \rightarrow \texttt{if (\textbf{b} == 0) return false; ...\textbf{a/b}...}$

%\item \texttt{Interprocedural Exceptions} \\
%$\texttt{...\textbf{bar()}...} \rightarrow$ \\ 
%$\texttt{try \{ \textbf{bar()} \} except (Exception e) \{ return false; \}...}$
%\end{enumerate} \vspace{.01cm}

\subsection{Removing Irrelevant Computation}
\label{sec:technique:sp2}
Lastly, we wish to remove program components which are unrelated to the precondition. This is an important step as it reduces cognitive load of reasoning over the precondition and removes unrelated statements which potentially have side effects. We consider a program segment to be \textit{unrelated} to the precondition if, when removed, the behavior of~\pre{} (modulo the test generator) does not change. We formulate this sub-problem as {\em program reduction} and leverage an off-the-shelf program reducer. A program reducer takes as input a program to reduce and a set of constraints. It then outputs the smallest sub-program which satisfies the given constraints. In our setting, the constraint is maintaining the behavior of~\pre{} (every generated test must pass). After the reduction phase, we have inferred a natural and correct \pre{}. We evaluate the impact of this reduction on the resulting preconditions in Section~\ref{sec:dataset:complexity}.

\subsection{Framework Instantiation Details}  In our framework, we instantiate the test generator as EvoSuite~\cite{EvoSuite} and leverage the Javaparser~\cite{javaparser} library to perform AST transformations. We instantiate the reducer as Perses \cite{perses}, as we experimented with the C-Reduce~\cite{c-reduce} program reducer but found it to be time-inefficient as it considers syntactically invalid sub-programs in the search space. 

\section{Comparative Case Study}
\label{sec:eval}

In this section, we perform an in-depth comparative evaluation to state-of-the-art approaches on a single real-world project. Here, we evaluate and compare the inferred preconditions on three aspects. We aim to answer the following research questions:

\begin{enumerate}
 \item[\textbf{RQ1}] Correctness: Are the inferred preconditions safe and maximal?
\item[\textbf{RQ2}] Naturalness: Can humans easily reason over the inferred preconditions?  
\item[\textbf{RQ3}] Model Affinity: How effective are inferred preconditions as few-shot examples in enhancing the performance of Large Language Models? 
\end{enumerate}

\subsection{Experimental Setup}
\textbf{Baselines.} To answer RQ1, we evaluate in comparison to \textsc{Proviso}\cite{pre-mod-test} as it is the state-of-the-art and instantiated in C\# which is most similar to our target language. We also compare to the popular invariant discovery technique, \textsc{Daikon}\cite{daikon}. As we will show in this section, \textsc{Daikon} often infers precoditions which are incorrect. As such, we forgo comparison to \textsc{Daikon} in answering RQ2 and RQ3. \\
%The other works most similar to ours, \textsc{PIE}\cite{pie} and \cite{gehr2015learning} are difficult to compare to as they are slightly different settings. \textsc{PIE} targets OCaml, however \textsc{Proviso}\cite{pre-mod-test} reports their search algorithm~\cite{id3, learning_com_specs} is comparable with a fixed feature set and has better performance when a feature set is not given. Thus we compare to the \textsc{Proviso}. \\
\textbf{Benchmark.} We evaluate on 39 $({\tt M}, \pre{})$ pairs from the \texttt{NetBigInteger} C\# project. Although our framework is designed for Java, we choose \texttt{NetBigInteger} as a common benchmark to compare to \textsc{Proviso}\cite{pre-mod-test}. In order to evaluate our technique, we manually translate the \texttt{NetBigInteger} class to a semantically equivalent class in Java. Although \textsc{Proviso} evaluates on 5 projects, 3 are synthetically constructed benchmarks, and 2 are real-world. We select a single project to evaluate on as manual effort is required in inspection and semantically equivalent conversion to Java. As such, we perform an in-depth evaluation on this single project. 

\subsection{RQ1: Correctness}
To measure correctness, we evaluate the safety and maximally of our inferred preconditions on the benchmark, modulo a test generator. To maintain a rigorous evaluation while balancing scalability, we evaluate correctness modulo the industrial symbolic exeutor, Pex~\cite{pex}. %By instantiating $T$ with a differing tool from the $T$ used in our technique (EvoSuite), we perform a higher quality evaluation as Pex has a fundamentally different input generation algorithm, and will likely produce differing inputs.  %but it can actually run at scale unlike a checker??? 
We invoke Pex on \method{} to generate crashing and non-crashing inputs. If Pex generates a crashing input that the precondition accepts, we consider the precondition to be \textit{unsafe}. If Pex generates a non-crashing input that the precondition rejects, we consider the precondition to be \textit{non-maximal}. If a precondition is both safe and maximal, we consider it to be correct. We perform this evaluation for the 39 preconditions inferred by each approach and report the results in Table~\ref{tab:rq1}.

\begin{table}[t]
\setlength{\belowcaptionskip}{1em}
\centering
\begin{tabular}{rccc|l}
	\toprule
	& Safe & Maximal & Correct & Total\\
	\midrule
	\textsc{Daikon}  & 25 & 10 & 6  & 39 \\
	\textsc{Proviso} & 37 & 34 & 34 & 39 \\
	\textsc{Ours}    & 34 & 32 & 32 & 39 \\
	\bottomrule
\end{tabular}
    \vspace{1em}
\caption{RQ1: Correctness of Inferred Preconditions.}
\label{tab:rq1}
\end{table}

We find that our inferred preconditions are safe for 34 methods, maximal for 32 methods, and both safe and maximal for 32 of the 39 methods. Through manual inspection, we find that the 7 incorrect \pre{} our approach infers are due to EvoSuite incompleteness. In comparison, \textsc{Proviso}'s preconditions are both safe and maximal for 34 methods, while \textsc{Daikon} achieves this for only 6 methods. This disparity is primarily because \textsc{Daikon} is designed as an invariant detector, discovering predicates true on all normally exiting runs rather than guarding against crashing runs. Additionally, \textsc{Daikon} is the only technique in this study with a passive algorithm, querying the test generator only once. In contrast, both our approach and \textsc{Proviso} actively query the test generator, terminating only when no further unguarded crashing inputs can be found. \\ 

\noindent \begin{longfbox}{\textbf{Result 1:} Our approach infers correct (both safe and maximal) preconditions for 32 of the 39 methods in the benchmark.}
\end{longfbox}

\subsection{RQ2: Naturalness} 
In this section, we evaluate the naturalness of our preconditions. To do so, we conduct a user evaluation to measure a human's ability to reason over the inferred preconditions. Each user is asked to review a given precondition and three inputs. We ask the user to classify each input as legal or illegal. The accuracy of their answers as well as the time taken to derive the answer are metrics of how readable or easy the precondition is to reason over.

\paragraph{User Study Design}
We identify 5 preconditions from our comparative evaluation to use filtered by the following criteria:
\begin{enumerate}
    \item Both approaches infer a correct precondition.
    \item The inferred preconditions are syntactically different.
    \item The precondition is non-trivial (there exists at least 1 illegal and 1 legal input). 
\end{enumerate}

\noindent We split participants into two groups. Group A was presented with our preconditions while Group B was presented with the preconditions inferred by \textsc{Proviso}. The inputs for each precondition remained the same across both groups. For illustration, 
we include one of the (precondition, input) pairs from our study. Figure~\ref{fig:p3a} shows an input given to Group A, while Figure~\ref{fig:p3b} shows one of the inputs given to Group B. We include any necessary code context for completing the task. For brevity, we only show a single input, but each user was presented with three inputs for each precondition. The user was asked to classify each input as True or False (legal or illegal). 

\begin{figure}
    \setlength{\belowcaptionskip}{1em}
    \begin{subfigure}[p]{.4\linewidth}
    \centering
    \lstset{style=mystyle, xleftmargin=.0\textwidth, xrightmargin=0.\textwidth, linewidth=0.0\linewidth}
\begin{lstlisting}[linewidth=3.4cm]
Input:

value: BigInteger                        
---------------------------
m_sign: int = -1                                       
m_magnitude: int[] =  {24}    
---------------------------

target: BigInteger
---------------------------
m_sign: int = 1
m_magnitude: int[] =  {10} 
\end{lstlisting}
    \label{fig:p3a:i2}
    \end{subfigure}
    \begin{subfigure}[p]{.4\linewidth}
    \lstset{style=mystyle, xleftmargin=.1\textwidth, xrightmargin=0.\textwidth, linewidth=0.0\linewidth}
\begin{lstlisting}[linewidth=4.5cm]
boolean M_pre(BigInteger value) {

  if (m_sign == 0)
     return true;

  if (value == null)
    return false;

  return true;


}
\end{lstlisting}
    \label{fig:p3:aours}
    \end{subfigure}
    %\vspace{-0.1in}
  \caption{Precondition 3, Input 2, Group A.}
  \label{fig:p3a}
\end{figure}

\begin{figure}
    \setlength{\belowcaptionskip}{1em}
    \begin{subfigure}[p]{.4\linewidth}
    \centering
    \lstset{style=mystyle, xleftmargin=.0\textwidth, xrightmargin=0.0\textwidth, linewidth=0.0\linewidth}
\begin{lstlisting}[linewidth=3.4cm]
Input:

value: BigInteger                        
---------------------------
m_sign: int = -1                                       
m_magnitude: int[] =  {24}    
---------------------------

target: BigInteger
---------------------------
m_sign: int = 1
m_magnitude: int[] =  {10} 
\end{lstlisting}
    %\hspace{.3cm}\includegraphics[width=5cm]{./user_study/P2/input1.png}
    \label{fig:p3b:i2}
    \end{subfigure}
    \begin{subfigure}[p]{.4\linewidth}
    \centering
    \lstset{style=mystyle, xleftmargin=.1\textwidth, xrightmargin=0.0\textwidth, linewidth=0.0\linewidth}
\begin{lstlisting}[linewidth=4.5cm]
(
 (value == null) 
  and (
      
      (not (target.m_sign <= -1)) 
      and ((target.m_sign <= 0))

  )

) or (
   (not (value == null))
)
\end{lstlisting}
    \label{fig:p3b:theirs}
    \end{subfigure}
    %\hfill
    %\begin{minipage}{\linewidth}
    %\vspace{2.2cm}
    %\begin{subfigure}[p]{\linewidth}
    %\lstset{style=mystyle, xleftmargin=.0\textwidth, xrightmargin=0.\textwidth, linewidth=0.0\linewidth}
    %\lstinputlisting[linewidth=7cm]{./examples/Add/Add_context.java}
    %\label{fig:p3b:context}
    %\end{subfigure}
    %\end{minipage}
    %\vspace{-0.1in}
  \caption{Precondition 3, Input 2, Group B.}
  \label{fig:p3b}
  %\vspace{-0.1in}
\end{figure}

\paragraph{Protocol}
The user study was conducted through a Qualtrics survey. We used fine grained timing for each participant. To reduce cognitive load, participants were randomly presented with three out of the five preconditions, using qualtrics randomization feature. Our survey included a ``Don't Know'' option to avoid random guessing. Any participant which selected ``Don't Know'' was not included in our results. The resulting valid participants comprised of 44 users including computer science PhD students, undergraduates, and industry software engineers split evenly between the two groups. 

\paragraph{User Study Results}
On average, we found that users were able to more accurately reason over our preconditions in a shorter amount of time. Participants of Group A had an average accuracy of 88.4\% while participants in Group B made more mistakes leading to a lower average accuracy of 79.71\% (p$<$.014). In terms of speed of reasoning, participants in Group A took an average total time of 150.89 seconds while participants in Group B took a longer amount of time to complete the study (204.98 seconds). Our results show that Precondition 1 (Figure~\ref{fig:mot_ex}) and Precondition 3 (Figures~\ref{fig:p3a} and~\ref{fig:p3b}) inferred by our approach were significantly easier to reason over in terms of both accuracy and speed (p$<$.0033). Precondition 4 (Figure~\ref{fig:P4}) had comparable results from both groups. However, on Precondition 2 (Figure~\ref{fig:P2}) and Precondition 5 results were weaker. %On Precondition 2, Group A achieved higher accuracy, but took more than 2x the average time as Group B (250.43 seconds vs 82.19 seconds). On Precondition 5, Group A achieved lower accuracy, but took a shorter amount of time than Group B. 
Preconditions 2 and 5 were the only samples in our study which included \texttt{try-catch} blocks, and thus, interprocedural reasoning. This indicates that although our preconditions are on average, easier to reason over, future research should address the additional cognitive load and lower level of readability due to interprocedural crashes. %We include all five preconditions from both groups A and B in this paper. Precondition 1 is shown in Figure~\ref{fig:mot_ex}, Precondition 3 is shown in Figures~\ref{fig:p3a} and~\ref{fig:p3b}, and we include Precondition 2 (Figure~\ref{fig:P2}), Precondition 4 (Figure~\ref{fig:P4}), and Precondition 5 (Figure~\ref{fig:P5}) in the appendix. \\

%\TODO{statistical significance????}

\noindent \begin{longfbox}{\textbf{Result 2:}  On average, users were able to more accurately reason over our preconditions in a shorter time span. Participants of Group A had an average accuracy of \textbf{88.84\%} while participants in Group B committed more errors, leading to a lower average accuracy of \textbf{79.71\%}.} %In terms of speed of reasoning, participants in Group A completed the task with an average total duration of \textbf{150.89 seconds}. Participants in Group B took a longer time (\textbf{204.98 seconds}) to finish the study. Our results were not as strong on preconditions which included interprocedural try-catch blocks.}
\end{longfbox} 
\vspace{1em}

\begin{table}[]
\setlength{\belowcaptionskip}{1em}
\begin{tabular}{l|ll|ll}
\toprule
      & \multicolumn{2}{c|}{Accuracy} & \multicolumn{2}{c}{Time Taken (Sec)} \\
        & Ours          & \textsc{Proviso}         & Ours          & \textsc{Proviso}         \\
    \hline\rule{0pt}{2.5ex}\texttt{Precondition 1}      &  97.78\%        &   77.78\%        &   92.23            &   309.49             \\
\texttt{Precondition 2}      &  80.56\%      &  82.05\%       &     250.43         &    82.19            \\
\texttt{Precondition 3}     &  100\%        &  73.33\%       &      104.14         &      216.76          \\
\texttt{Precondition 4}      &  97.22\%        &  84.85\%         &     61.77          &     68.38          \\
\texttt{Precondition 5}      &  68.63\%        &  80.56\%       &      245.88         &     348.09           \\ 
\hline\rule{0pt}{2.5ex}Overall &    \textbf{88.84\%}           &  79.71\%              &  \textbf{150.89}             &   204.98            \\
\bottomrule
\end{tabular}
\label{tab:user_study}
\vspace{1em}
\caption{User study results.}
%\vspace{-0.4in}
\end{table}

%\paragraph{Naturalness of \textsc{Daikon}'s inferred preconditions}
%Given \textsc{Daikon}'s relatively low level of correctness, we focused our comparison solely on \textsc{Proviso} in this section. However, we note that, even when correct, \textsc{Daikon}'s inferred preconditions contain many irrelevant variables. \TODO{quantify?.}

\subsection{RQ3: Model Affinity} 
In this section, we evaluate the affinity of statistical models of code toward the inferred preconditions. To measure this, we evaluate GPT-4's performance on inferring preconditions for the \texttt{NetBigInteger} benchmark in three settings: zero shot, few shot with (\method{}, \pre{}) pairs, and few shot with (\method{}, proviso precondition) pairs. To create the prompt, we select five diverse samples from our dataset. We report the results in Table~\ref{tab:rq3}. We find that without any explicit training examples in the prompt, GPT-4 was able to infer correct precondtions for 21 out of the 39 samples. When supplied with five (method, precondition) pairs generated by \textsc{Proviso}, GPT-4's output slightly improved as it was able to generate 2 additional correct preconditions. When supplied with the same five methods, but with \pre{}'s generated by our approach the quality of preonditions rose and GPT-4 was able to generate 28 correct preconditions. We note that GPT-4, even without any few-shot examples, performed significantly better than \textsc{Daikon}. This result indicates future research in neural precondition generation as a promising direction.

\begin{table}[t]
\setlength{\belowcaptionskip}{1em}
\centering
\begin{tabular}{rccc|l}
	\toprule
	& Safe & Maximal & Correct & Total\\
	\midrule
	Zero Shot                   & 23 & 32 & 21  & 39 \\
    Few Shot - \textsc{Proviso} & 28 & 25 & 23 & 39 \\
    Few Shot - \textsc{Ours}    & 33 & 34 & 28 & 39 \\
	\bottomrule
\end{tabular}
\vspace{1em}
\caption{RQ3: Correctness of GPT-4s Inferred Preconditions.}
\label{tab:rq3}
\end{table}

Beyond correctness, we found that precondition quality improved when supplied with few shot prompts generated by our approach in terms of size and number of irrelevant clauses. We consider a clause to be irrelevant if it can be removed without impacting the correctness of the precondition. Of the samples it inferred correctly, GPT-4 Zero Shot produced extraneous clauses in 24\% of the cases. In contrast, when supplied with our preconditions in the prompt, GPT-4 only produced irrelevant or redundant clauses in 18\%. However, when supplied with \textsc{Proviso} generated preconditions, the quality dropped and extraneous clauses were included in 35\% of the correctly inferred preconditions.

%GPT also ran much faster and gives a good insight that future work could be neural. Performed better than Daikon!!
\vspace{2mm}
\noindent \begin{longfbox}{\textbf{Result 3:} GPT-4 was able to correctly infer 28 out of 39 preconditions when supplied with few shot examples from our dataset.}
\end{longfbox}

\section{A Dataset of Preconditions}
%\section{Large Scale Real-World Evaluation}
\label{sec:dataset}
In this section, we present a dataset of real-world Java preconditions and its characteristics. We use the dataset to answer the following research questions: 

%In particular, we evaluate the impact of active queries to test generator and the impact of program reduction. To do so, we gather a substantial dataset of real-world Java preconditions and answer the following research questions: 
%We present the implementation of our approach as a tool and demonstrate its application on a significant scale across 87 real-world Java projects. This yields $\sim$18k (method, precondition) pairs, obtaing the first large-scale dataset of preconditions. This dataset not only underscores the viability and effectiveness of applying our tool at scale, but also provides a benchmark for evaluation. In this paper, we use this benchmark to conduct a large-scale evaluation of the design decisions made to enhance effectiveness of our approach. We also present characteristics of the dataset which in turn offers valuable insights to guide future research efforts in the field of precondition inference.
%We aim to evaluate our technique's design decisions in both sub-problems. First, we evaluate our choice of iterative, dynamic, exception check insertion over the two-step approach presented in Section~\ref{sec:technique:naive:two}.
%It is difficult to evaluate exhaustive up-front transformation approach presented in Section~\ref{sec:technique:naive:ex} since our definition of correctness is modulo a test generator. For this, we default to an illustrative example by contradiction of correctness in Figure~\ref{fig:naive}. 
%Secondly, we motivate and evaluate our solution to removing irrelevant program segments in terms of complexity. 

\begin{enumerate}
\item[\textbf{RQ1}] How does active test generation impact the quality of test suite and resulting precondition?
\item[\textbf{RQ2}] How does removing irrelevant program segments affect the resulting precondition in terms of complexity? 
\end{enumerate}

%\subsection{Experimental Setup: Benchmark}
%Our dataset is constructed of Java methods from popular, stable, projects. 
Our dataset consists of 17,871 $(\method{}, \pre{})$ pairs inferred from 87 projects. The projects are from the SF100~\cite{EvoSuite-challenges} with an additional 13 projects popular projects from the Maven repository~\footnote{The projects were collected in order of number of downloads as of April 2022.}. In order to provide further insight into our benchmark, we present the following characteristics of our dataset:

\vspace{1em}
\begin{table}[ht]
\centering
%\begin{subfigure}{.43\textwidth}
%\centering
%{\small
%\begin{tabular}{l|rr}
%\toprule
%            & \multicolumn{1}{l|}{Number} & Percent \\ \hline
%non-trivial & \multicolumn{1}{l|}{6,790}  & 37.99\% \\
%true        & \multicolumn{1}{l|}{10,791} & 60.38\% \\
%false       & \multicolumn{1}{l|}{290}    & 1.62\%  \\ \hline
%Total       & \multicolumn{2}{l}{17,871}           \\ 
%\bottomrule
%\end{tabular}}
%\vspace{0.1in}
%\caption{Overall Precondition Categorization.}
%\label{tab:dataset}
%\end{subfigure}
%\vspace{0.2in}
%\begin{subfigure}{.55\textwidth}
%    \begin{figure}{\textwidth}
%    \begin{table}
%\centering
\setlength{\belowcaptionskip}{1em}
%{\small
	\begin{tabular}{c|r}
	\toprule
        Crash Type                              & \# of Checks  \\
        \cmidrule(l){1-2}
        Intraprocedural Exception               &  8,079     \\
		\texttt{NullPointerException}           &  5,535    \\
  		\texttt{ArrayIndexOutOfBoundsException} &  793     \\
    	\texttt{ClassCastException}             &   443    \\
		\texttt{NegativeArraySizeException}     &   65    \\
		\texttt{ArithmeticException}            &   17     \\
          \cmidrule(l){1-2}
		Overall                                 & 14,932 \\
	 \bottomrule
	\end{tabular}
%    }
    \vspace{1em}
%\vspace{0.1in}
%\caption{Breakdown by Crash Type.}
%\end{subfigure}
\caption{Dataset Crash Type Statistics.}
\label{tab:data-breakdown}
%\vspace{-0.2in}
\end{table}
%\vspace{-.2in}

\paragraph{Crash Occurrence} Since our dataset is real-world and diverse, it contains methods which accept all inputs, reject all inputs, or have a non-trivial precondition which accepts some inputs and rejects others. We characterize the dataset, by the precondition "type". In our dataset, more than half of the methods (6,790) have a precondition of \texttt{true} indicating that any input is legal. A small percentage ($\sim$1\%) of samples (290) are always exceptional and have a precondition of \texttt{false}. Through manual inspection, we find these methods typically contain a single explicit \texttt{throws} expression. For example, a method with a body of: \texttt{throw new UnsupportedOperationException} would always be exceptional and have a precondition of \texttt{false}. The remaining samples, (10,791), are non-trivial in that there exists at least one legal input and one illegal input. 

%textcolor{red}{In the C\# Benchmark, 40\% of methods had a precondition of true}.
\paragraph{Crash Types} EvoSuite found a diverse set of crashes which fit into the 6 crash type categories presented in Section~\ref{sec:technique:sp1}. In Table~\ref{tab:data-breakdown}, we present the total number of explicit exception checks inserted and the occurrence of each type. These checks were dynamically lifted to explicit guards by the Check Instrumentor and were not present in the source of \method{}. Nearly 15,000 checks were added through program transformation. This emphasizes the need for a dynamic tester to find crashing inputs rather than being able to parse out explicit checks as in Figure~\ref{fig:mot_ex} line 3. We also note that over half of crashes found by EvoSuite occurred in a callee, indicating a potential fruitful research direction in natural precondition inference.

\paragraph{Defensive Checks} Here, we ask, how often do developers explicitly guard for precondition violations? This is in converse to the previous characterization in that we study only the checks which were included in the untransformed ~\method{}. Through this study, we motivate the necessity of a dynamic tester to find legal and illegal inputs beyond the explicitly written exceptions. Previous approaches have attempted to statically create such a dataset of preconditions through parsing. One static parsing approach~\cite{bodyguard} attempts to generate a dataset of invariants. The authors do this by considering an if-statement's guard condition to be an invariant of the guarded block. Although they were able to collect these invariants on a large scale (2.5 million invariants), such a dataset cannot capture implicit specifications that are not expressed directly in \texttt{if} conditions. Although our setting is slightly different in that we are collecting preconditions rather than invariants, the approach still stands as a naive evaluation. 

Through an AST walk over the target methods in our dataset we find 1,451 explicitly coded illegal input guards. This, in conjunction with the nearly 15,000 implicit crashes found shows that high quality preconditions cannot be constructed by simple scraping of if conditions. It also indicates that developers do not sufficiently guard against illegal inputs \\

%\noindent \begin{longfbox}{We find that only \textbf{1,451} guards were explicitly written by developers. The remaining \textbf{14,932} crashes were not explicitly guarded against and were found by EvoSuite.}
%\end{longfbox} 

%\paragraph{Qualitative Study} \textcolor{red}{We will also include qualitative examples}
 
%\subsection{Large Scale Evaluation.}
%Here, we aim to answer the research questions using the dataset presented above. 
\subsection{RQ1: Impact of Active Test Generation} 
Here, we empirically motivate the need for active queries to the test generator on a large scale. To do so, we measure the number of samples in our dataset which required iteration. That is, the number of samples for which EvoSuite finds new crashing inputs when it is invoked on $M\_pre\_1$ which includes inserted checks. 
We find that 2,198 samples required additional iterations beyond a single step process. This amounts to 32.37\% of our non-trivial samples. 
During the inference of these 2k samples, they required on average, 1.73 additional iterations (beyond the initial step). In total, 4,517 additional crashes were found due to additional iterations. \\

\noindent \begin{longfbox}{\textbf{Result 1:} We find that over 1/3rd of non-trivial samples require additional iterations beyond a single step (test and transform) approach. A total of 4,517 additional crashes were found due to the iterative approach.}
\end{longfbox} 
\subsection{RQ2: Impact of Program Reduction} 
\label{sec:dataset:complexity}
In this section, we evaluate the effect of removing irrelevant computation. The goal of removing irrelevant program segments is to reduce cognitive load and remove unrelated statements which potentially have side effects. In short, program reduction should reduce complexity. We evaluate this along three axis: number of program paths, precondition size, and purity. For each axis, we compare the unreduced \pre{} (post-instrumentation) to the reduced \pre{}.

\paragraph{Cyclometic Complexity}
In order to quantify the complexity of a precondition's control flow, we measure the cyclomatic complexity~\cite{cyclocomplexity} using checkstyle~\cite{checkstyle}. The average cyclometic complexity of an unreduced sample is 2.86. After reduction, the complexity drops to 1.77. This shows the computational load is reduced by around one path through program reduction. For non-trivial samples with at least one illegal and one legal input, the average complexity is 4.99 and drops to 2.46 after reduction. For this sub-class of samples, complexity is more significantly lowered through reduction. 

\paragraph{Precondition Size}
In order to quantify a precondition's size, we measure the number of AST nodes. The average number of unreduced nodes is 56.27 and the average number of AST nodes in the reduced sample is 15.41 This shows that we removed around 1/3 of the AST nodes through program reduction, greatly reducing computational load of a consumer of the precondition. 

\paragraph{Purity}
Ideally, a precondition should not have side effects. Impure statements increase cognitive load, forcing the consumer to reason about state, and add complications for precondition execution during software testing. We measure the number of samples which are impure pre-reduction, but pure post-reduction to quantify impact of program reduction on side effects. To do so, we use the Facebook Infer~\cite{fb-infer} purity checker. We find that only 50.63\% of pre-reduction samples are pure, while 81.3\% post-reduction samples do not contain side effects. This demonstrates the impact of reduction on the purity of inferred preconditions.  \\

\noindent \begin{longfbox}{\textbf{Result 2:} We find that by employing a program reducer we produce preconditions which are of lower complexity, smaller, and display a higher level of purity.} %Cyclometic complexity drops by around 1 program path after reduction. For non-trivial samples, this is more significant, with reduction lowering complexity by half. The number of AST nodes is reduced by 1/3rd. Post reduction, there is a 60\% increase in number of pure samples.} %This motivates the use of program reduction across three axis of complexity.}
\end{longfbox} 
%Through the presentation of this dataset, we also demonstrate the potential for large-scale application of our approach and practicality of our open source tool. In addition to a large scale evaluation of our approach, we analyze the dataset to gain valuable insights that can guide and stimulate further research in precondition inference. 

\section{Limitations}
\label{sec:limitations}

Our approach, like many other precondition inference techniques, is limited by program analysis capabilities. Firstly, our approach heavily relies on the target method implementation.
If the source code is not available, or does not compile and execute, then our technique cannot infer a precondition.
Likewise, if the method has a bug related to the precondition, it may be inherited in the resulting~\pre{}.

Secondly, while instantiating our technique as a tool, we found that program analysis tools are very brittle. We applied our technique at scale on a total of 126,699 methods and only obtained 17,868 $({\tt M}, \pre{})$ pairs ($\sim$14\% of the target methods). The methods we could not extract preconditions for failed due to limitations of the program analysis tools in our pipeline. Failures largely occurred due to automated testing challenges in EvoSuite. For example, EvoSuite often could not generate a well-formed object as a target for {\tt M}. EvoSuite struggled to generate test suites in some cases due to resource limits, I/O and file system mocking, concurrency, and missing environment dependencies~\cite{EvoSuite-challenges}. 

A limitation unique to our approach is the possibility of side effects. Since our approach relies on the structure and program segments that appear in the original ~\method{}, there is no guarantee that the precondition we produce is pure. In practice, we find that most preconditions infer are pure in nature. Although, if purity is critical, a purity checker such as FBInfer~\cite{fb-infer} can be invoked to avoid the rare case where the method's precondition has side effects.

%Lastly, the inherent nature of program analysis limits us to an instantiation in a single language (Java). 
%\vspace{-9cm}
% In tool usage, a precondition with side effects must be run in a sandbox environment as to preserve the state of the program being analyzed. The challenge of maintaining purity is unique to our approach. Prior works have the benefit of ensuring purity by construction. That is, their precondition features are designed to only include observer methods which do not change the program state.
%\TODO{Paper ending looks abrupt; add a Conclusion section.  Also avoid large chunks of empty space even though it is Latex's fault, by using -ve vspace construct or adjucting placement of figures.}
%Lastly, we note that our approach performed weakest on inferring preconditions for methods with interprocedural crashes 
%\input{8_future_work.tex}
%\vspace{-9cm}
\section{Conclusion}
\label{sec:conclusion}

We have presented a novel approach for inferring  natural preconditions through program transformations over the target method. Our findings reveal that the preconditions inferred by our approach are more easily reasoned over by humans than template based tools which build the precondition from scratch. Looking ahead, the potential of natural preconditions extends to diverse applications such as serving as training data or integrating with large language models. This exciting direction is underscored by the promising advancements in the realm of program specification representations as evidenced in \cite{sutton-signatures}. %along with the high accuracy demonstrated by large language models in inferring Daikon invariants, as showcased in \cite{sutton-daikon}. 
The concept of natural preconditions is in alignment with the evolving landscape of statistical software engineering tools, especially within the domain of large language models. As this synergy continues to unfold, we anticipate implications for the future of software correctness.

\pagebreak

\bibliographystyle{ACM-Reference-Format}
\bibliography{refs}
%\pagebreak
\newpage
\section{Appendix}
\label{sec:app}

\begin{figure}[htbp]
\setlength{\belowcaptionskip}{1em}
\centering
\begin{subfigure}{.4\linewidth}
    \lstset{style=mystyle, xleftmargin=.0\textwidth, xrightmargin=0.\textwidth, linewidth=0.0\linewidth}
\begin{lstlisting}[linewidth=4.5cm]
boolean M_pre(NetBigInteger value) {
  try {
    CompareTo(value);
  } catch (NullPointerException e) {
    return false;
  }
  return true;
}
\end{lstlisting}
    \caption{Group A.}
    \label{fig:P2:B}
    \end{subfigure} 
    \begin{subfigure}{.4\linewidth}
    \lstset{style=mystyle, xleftmargin=.1\textwidth, xrightmargin=0.\textwidth, linewidth=0.0\linewidth}
\begin{lstlisting}[linewidth=3.1cm]
((not (value == null)))
\end{lstlisting}
    \caption{Group B.}
    \label{fig:P2:B}
    \end{subfigure} 
\caption{Precondition 2}
\label{fig:P2}
\end{figure}

\begin{figure}[htbp]
    \setlength{\belowcaptionskip}{1em}
    \centering
    \begin{subfigure}{.4\linewidth}
    \lstset{style=mystyle, xleftmargin=.0\textwidth, xrightmargin=0.\textwidth, linewidth=0.0\linewidth}
\begin{lstlisting}[linewidth=4.0cm]
boolean M_pre(NetBigInteger m) {
  if (m == null)
    return false;
  if (m.m_sign < 1)
      return false;
  return true;
}
\end{lstlisting}
    \label{fig:P4:A}
    \caption{Group A.}
    \end{subfigure}
    \begin{subfigure}{.4\linewidth}
    \lstset{style=mystyle, xleftmargin=.1\textwidth, xrightmargin=0.\textwidth, linewidth=0.0\linewidth}
\begin{lstlisting}[linewidth=4.0cm]
(not (m == null)) and
(not (m.m_sign <= 0))
\end{lstlisting}
      \caption{Group B.}
     \label{fig:P4:B}
    \end{subfigure}
    \caption{Precondition 4.}
  \label{fig:P4}
\end{figure}
%
%   PRECONDITION 5
%
\begin{figure}[htbp]
\centering
\setlength{\belowcaptionskip}{1em}
    \begin{subfigure}{\linewidth}
    \lstset{style=mystyle, xleftmargin=.1\textwidth, xrightmargin=0.0\textwidth, linewidth=0.0\linewidth}
\begin{lstlisting}[linewidth=6.7cm]
boolean M_pre(NetBigInteger value) {
  if (value == null)
    return false;
  if (m_sign == 0)
    return true;
  NetBigInteger r;
  NetBigInteger u = this;
  NetBigInteger v = value;
  while (v.m_sign != 0) {
    try {
      r = u.Mod(v);
    } catch (ArithmeticException e) {
      return false;
    }
    v = r;
  }
  return true;
}
\end{lstlisting}
    \caption{Group A.}
    \label{fig:P5}
    \end{subfigure}
    \begin{subfigure}{\linewidth}
    \lstset{style=mystyle, xleftmargin=.0\textwidth, xrightmargin=0.\textwidth, linewidth=0.0\linewidth}
\begin{lstlisting}[linewidth=8cm]
((not (value == null)) and (not (target == null))) 
and ( 
      (-target.m_sign + -value.m_sign <= 1) and (
        (not (-target.m_sign + value.m_sign <= -2))
      )
)
\end{lstlisting}
    \caption{Group B.}
    \label{fig:p5:B}
    \end{subfigure}
  \caption{Precondition 5.}
  \label{fig:P5}
\end{figure}

% GROUP B

%\begin{figure}[!htb]
%    \begin{subfigure}[p]{\linewidth}
%    \lstset{style=mystyle, xleftmargin=.0\textwidth, xrightmargin=0.\textwidth, linewidth=0.0\linewidth}
%    \lstinputlisting[linewidth=8cm]{./examples/GCD/GCD_theirs.java}
%    \label{fig:p1:ours}
%    \end{subfigure}
%    
%  \caption{Precondition 5, Group B.}
%  \label{fig:P5B}
%\end{figure}

\end{document}